\documentclass[aps,twocolumn,showpacs,preprintnumbers,prl,a4paper,superscriptaddress]{revtex4}
\usepackage{graphicx}
\usepackage{amssymb}
\usepackage{amsmath}
\usepackage{verbatim}

\def\dens{\rho}
\def\adens{\tilde{\dens}}
\def\etau{\tau^{\text{KS}}}
\def\taup{\tau^{\text{p}}}
\def\ataup{\tilde{\tau}^{\text{p}}}
\def\tautf{\tau^{\text{TF}}}

\usepackage{color}

\begin{document}

\title{Kohn-Sham Kinetic Energy Density in the Nuclear and Asymptotic Regions: 
Deviations from the Von  Weizs\"{a}cker Behavior and Applications to Density Functionals}
\author{Fabio Della Sala}
\affiliation{National Nanotechnology Laboratory (NNL), Istituto Nanoscienze-CNR,
Via per Arnesano 16, I-73100 Lecce, Italy }
\affiliation{Center for Biomolecular Nanotechnologies @UNILE, Istituto Italiano di Tecnologia,
Via Barsanti, I-73010 Arnesano, Italy}
\date{\today}
\author{Eduardo Fabiano}
\affiliation{National Nanotechnology Laboratory (NNL), Istituto Nanoscienze-CNR,
Via per Arnesano 16, I-73100 Lecce, Italy }
\affiliation{Center for Biomolecular Nanotechnologies @UNILE, Istituto Italiano di Tecnologia,
Via Barsanti, I-73010 Arnesano, Italy}
\author{Lucian A. Constantin}
\affiliation{Center for Biomolecular Nanotechnologies @UNILE, Istituto Italiano di Tecnologia,
Via Barsanti, I-73010 Arnesano, Italy}

\begin{abstract}
We show that the Kohn-Sham positive-definite kinetic energy (KE) density significantly differs from the  
von Weizs\"{a}cker (VW) one at the nuclear cusp as well as in the asymptotic region.
At the nuclear cusp, the VW functional is shown to be linear and the contribution of $p$-type orbitals to the KE density 
is theoretically derived and numerically demonstrated in the limit of infinite nuclear charge,
as well in the semiclassical limit of neutral large atoms. 
In the latter case, it reaches $12\%$ of the KE density.  
In the asymptotic region we find new exact constraints for meta 
Generalized Gradient Approximation (meta-GGA) exchange functionals: 
with an exchange enhancement factor proportional to $\sqrt{\alpha}$, 
where $\alpha$ is the common meta-GGA ingredient, 
both the exchange energy density and the potential are proportional to 
the exact ones. In addition, this describes exactly the large-gradient limit
of quasi-two dimensional systems.   
\end{abstract}


\maketitle

Ground-state density functional theory (DFT)  \cite{HK,KS,dftbook,dftbook2}
can be considered the most used method in electronic
calculations of quantum chemistry and condensed matter
physics. DFT is in principle an exact approach to electronic structure theory.
However, in practice the exchange-correlation (XC)
and, in the case of orbital-free DFT \cite{of}, the non-interacting kinetic energy (KE)
functionals  need to be approximated.
The construction of approximate XC and KE functionals is thus
an active field within DFT \cite{xcrev,kinrev,ofrev}.

The Kohn-Sham (KS) positive-definite KE density $\etau=(1/2)\sum_i|\nabla\phi_i|^2$, 
where $\phi_i$ are the occupied KS orbitals, is of course a key
quantity for KE functionals but plays a major role also in 
meta-Generalized Gradient Approximation (meta-GGA) XC energy functionals, 
which are recently attracting strong interest \cite{Vignale,revTPSS,BLOC,sun13,m11l}.
The properties of $\etau$ have been studied since long \cite{dftbook,dftbook2},
suggesting that it approaches the von Weizs\"{a}cker (VW) kinetic 
energy density functional \cite{vW}, $\tau^W[\dens]=|\nabla \dens|^2/(8\dens)$,  
both in the cusp \cite{dftbook2,burke,kara,bader72,alva,lastra} and in the asymptotic regions 
\cite{dftbook2,burke,alva,vitos,ernz00,llMGGA}.
These conditions have been used in the derivation of approximated KE functionals \cite{lastra,vitos,ernz00,llMGGA}.

Concerning the nuclear region, it was more recently shown that $\etau$ differs from 
$\tau^W$ at the nuclear cusp \cite{quian}.
In this work we will present a new derivation of the KE at the nuclear cusp which is based on the 
linearity of the VW functional at the cusp.
Moreover, we will derive the semiclassical limit for the exact KE density as well as the limit for isoelectronic series
at the  nuclear cusp.

Concerning the asymptotic region of atoms it has been previously observed \cite{ernzer,elf,johnson09} that  
$\etau$ can have different asymptotic properties from $\tau^W$, if the outer-valence electrons are not of  $s$-type. 
In this work, we will show an exact asymptotic expression for 
the Pauli excess KE density \cite{dftbook,kara,levy}
$\taup=\etau-\tau^W$, and use it to derive a new constraint
for the asymptotic behavior of meta-GGA exchange functionals.
This is an almost unexplored topic since, so far, most of the effort has been put 
into the investigation of the asymptotic properties of GGA functionals.
However, the latter cannot display simultaneously the correct asymptotic
behavior for both the exchange energy per particle ($\epsilon_x\rightarrow-1/(2r)$) and 
the exchange potential $v_x\rightarrow-1/r$ \cite{noasy,arm13}.
Thus, e.g., the Becke exchange \cite{becke} has an exact asymptotic behavior for
$\epsilon_x$, but $v_x$ is proportional to $-1/r^2$ \cite{becker2}. On the other hand, 
a recent functional \cite{arm13} with the correct $v_x\rightarrow-1/r$ has been developed, 
but it has a nonphysical $\epsilon_x$. 
In this letter we will show instead that the aforementioned constraints
can be merged at the meta-GGA level. 

We start considering a closed-shell system in an arbitrary central 
spherical potential $V(r)$ (e.g. atoms, jellium spheres, ...).
A given shell is characterized by $n,l$ quantum number 
and all orbitals with $m=-l,\ldots,l$  
contribute to  shell density $\dens_{nl}$.
The positive-definite KE density of the shell can be written as \cite{suppinfo,nagymarch}:
\begin{equation}
\label{eq5}
\etau_{nl}  = \tau^W[\dens_{nl}] +\frac{l(l+1)}{2} \frac{\dens_{nl}}{r^2} \, .
\end{equation}
The total KS positive-defined KE density is
$\etau=\sum_{nl} \etau_{nl}$. Note that this simple formula
does not apply to the non-linear VW functional,
since $\tau^W[\dens] \neq \sum_{nl} \tau^W[\dens_{nl}]$.
Eq. (\ref{eq5}) is the starting equation for this work and it 
is valid \emph{everywhere} in the space, for any shell and 
any central potential. In the following we consider two limits for 
Eq. (\ref{eq5}), namely $r\rightarrow 0$ and 
$r\rightarrow \infty$.

When $r\rightarrow 0$  we have that (for $l\ge 1$) 
$\dens_{nl} \rightarrow A_{nl} r^{2l}$, where $A_{nl}$ 
is a constant. Thus, the second term on the right-hand-side of 
Eq. (\ref{eq5}) is $A_{n1}$ for $l=1$ and vanishes for $l\ge2$. 
For the VW term at $r=0$ we have for $l=1$
\begin{eqnarray}
\tau^W[\dens_{n1}](0)  =  \frac{1}{8}\frac{(2A_{n1}r)^{2}}{A_{n1}r^{2}}=\frac{1}{2}A_{n1} \, ,
\label{eq:kpt}
\end{eqnarray}
whereas $\tau^W[\dens_{n,l}](0)$ vanishes for $l \ge 2$ and 
for $l=0$  it depends on the central potential.
Summarizing, we have
\begin{equation}
\label{eqt}
\etau_{nl}(0)=\left \{
\begin{array}{ll}
\tau^W[\dens_{n0}](0)         & \mathrm{for} \;\; l=0 \\
\frac{1}{2} A_{n1}+A_{n1}=3 \tau^W[\dens_{n1}](0)       & \mathrm{for} \;\; l=1\\ 
0                            & \mathrm{for} \;\; l\ge 2 \\
\end{array}
\right . 
\, .
\end{equation} 
%

To stress the significance of this result we consider the special case of a central potential with a leading term  $-Z/r$ near the core. 
In this case the Kato theorem \cite{kato}
holds for any spherical shell \cite{esquivel}, i.e. 
$\left. \frac{\partial \rho_{n0}(r)}{\partial r} \right|_{r=0}=-2Z\rho_{n0}(0)$.
Therefore
\begin{equation}
\label{eq:kst}
\tau^W[\dens_{n0}](0) = \frac{1}{8}\frac{\left[-2 Z \dens_{n0}(0)\right]^2}{\dens_{n0}(0)}=\frac{Z^2}{2}\dens_{n0}(0) \, .
\end{equation}

\begin{figure}[hbt]
\includegraphics[width=\columnwidth]{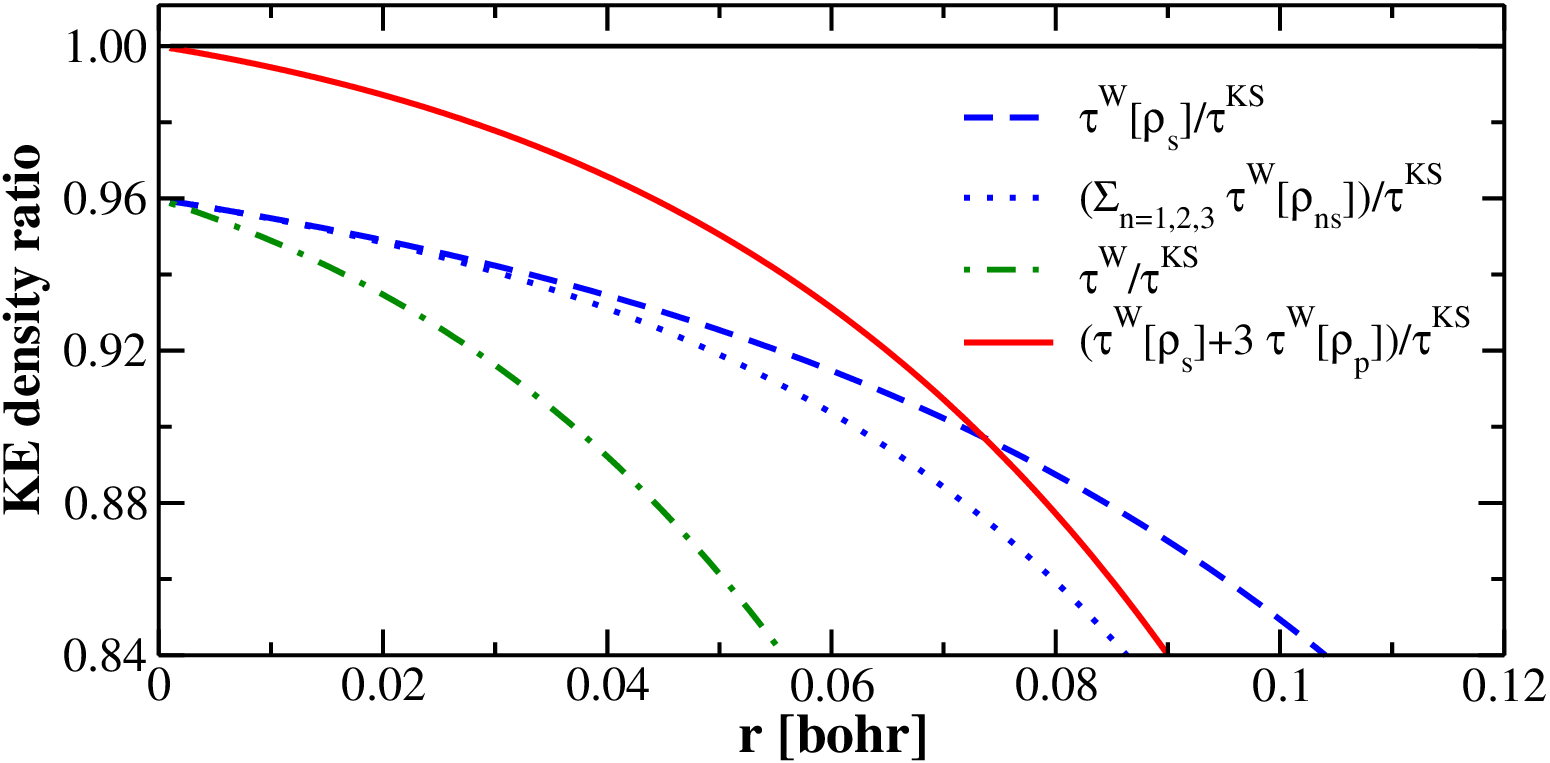}
\caption{\label{figatt}Ratios of different kinetic energy (KE) densities to the exact KE ($\etau$) 
for the Argon atom, near the nucleus. 
The dashed and dotted lines coincide at $r=0$, showing the linearity of the VW functional for $s$-shells. 
The red solid line approaches 1 at $r=0$ showing the validity of Eq. (7).}
\end{figure}

The main finding here is that Eqs. (\ref{eq:kst}) and (\ref{eq:kpt}) are {\it linear} 
in $\dens_{n0}(0)$ and $A_{n1}$, respectively. Thus  we have:
\begin{eqnarray}
\tau^{W}[\dens_{s}](r) & = & \sum_{n=1}^{N_s} \tau^W[\dens_{n0}](r)  \;\;\; {\rm for} \;\;\; r=0  \, ,
\label{lltt1} \\
\tau^W[\dens_{p}](r) & = & \sum_{n=2}^{N_p} \tau^W[\dens_{n1}](r)   \;\;\; {\rm for} \;\;\; r=0 
\label{lltt2} \, ,
\end{eqnarray}
where $\dens_s=\sum_{n=1}^{N_s} \dens_{n0}$ and $\dens_p=\sum_{n=2}^{N_p} \dens_{n1}$, 
with $N_s$ and $N_p$ being the maximum principal quantum numbers for occupied 
$s$-type and $p$-type shells respectively.
We underline that the linearity of the VW functional is valid only at the nuclear position.
This is shown in Fig. \ref{figatt} where both sides of Eq. (\ref{lltt1}) are reported versus the radial distance  
for, e.g., the Argon atom. For the corresponding plot of both sides of Eq. (\ref{lltt2}), see Ref. \cite{suppinfo}. 
All calculations in this work have been performed with a numerical atomic code using non-relativistic exact-exchange (EXX)\cite{APBE}.

Combining Eqs. (\ref{lltt1}) and (\ref{lltt2}) with Eq. (\ref{eqt}) we have for 
the total positive-defined KE density at the origin:
\begin{eqnarray}
\etau (0)  &=& \sum_{nl} \etau_{nl}(0)
=\tau^{W}[\dens_{s}](0)+ 3\tau^W[\dens_{p}](0) \label{ext} \, .
\end{eqnarray}
Hence, the exact positive-defined KE density at the origin depends
on both $s$-type and $p$-type contributions.
A numerical evidence of the validity of Eq. (\ref{ext}) for, e.g., the Argon atom is given in Fig. \ref{figatt}.

\begin{figure}[hbt]
\includegraphics[width=\columnwidth]{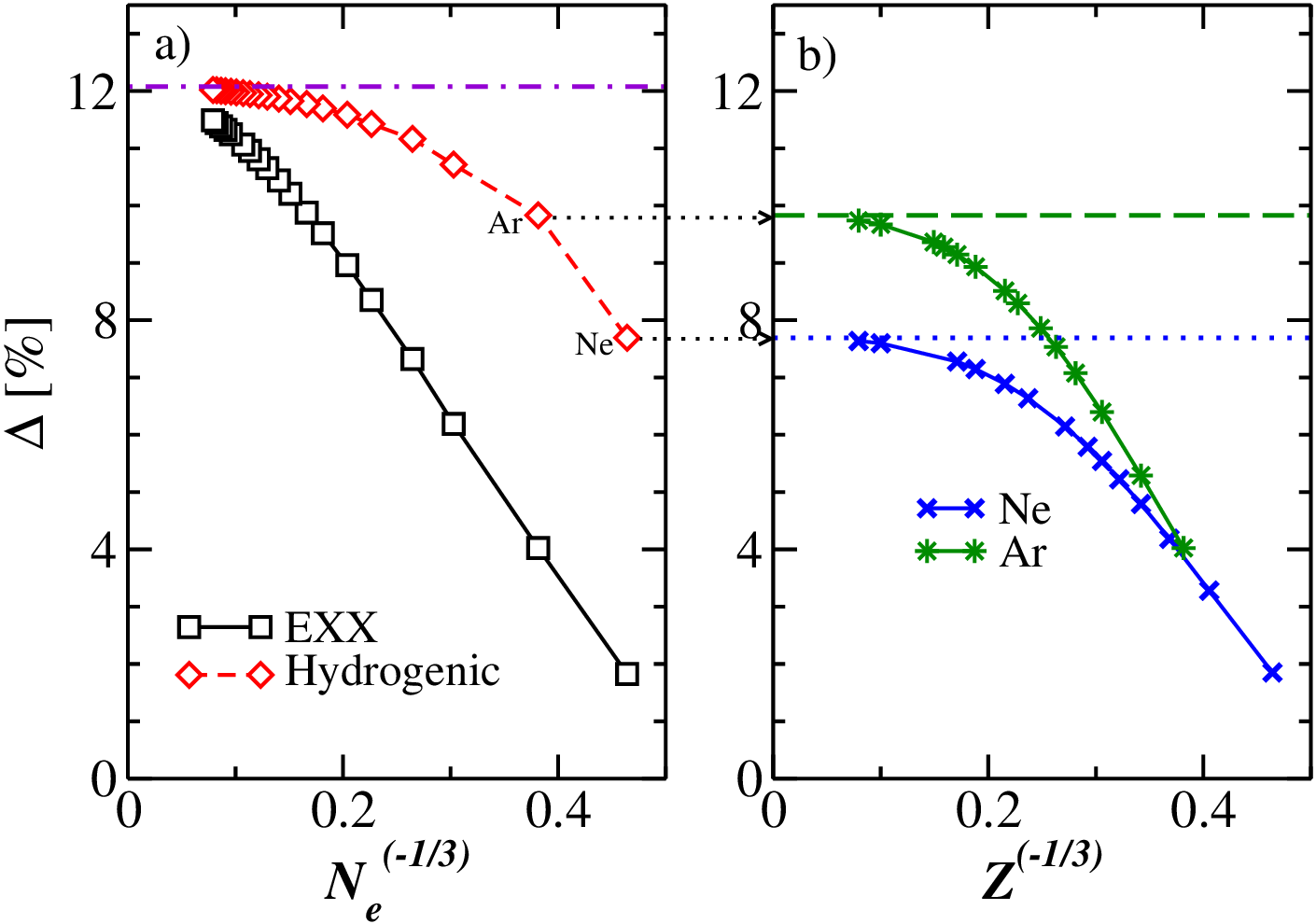}
\caption{\label{figpp}
Relative contribution ($\Delta$ in percent) of $p$-shells to the positive-defined KS kinetic energy density 
at the nuclear cusp for:
a) neutral noble-gas atoms up to $N_e$=2022 as a function of  $N_e^{-1/3}$. Results using EXX 
and hydrogenic orbitals are shown.
The horizontal dashed-line represents the theoretical limit, see Eq. (11).
b) Ne and Ar isoelectronic series with EXX orbitals up to $Z=2000$. 
The horizontal dashed-lines represent the theoretical limits, see Eq. (10).
}
\end{figure}

In order to estimate the effective role of the $p$-type 
orbitals on real calculations we consider all neutral closed-shell noble 
atoms up to $N_e$=Z=2022 (i.e. up to 21 $p$-shells; here and hereafter $N_e$ is the number of 
electrons and $Z$ is the nuclear charge).
In Fig. \ref{figpp}a we report the quantity
\begin{equation}
\Delta^\mathrm{EXX}= \frac{3\tau^W[\dens_{p}](0)}{\tau (0)}
=  \frac{3\tau^W[\dens_{p}](0)}{\tau^{W}[\dens_{s}](0)+ 3\tau^W[\dens_{p}](0)}
\, ,
\end{equation}
as a function of ${N_e}^{-1/3}$.
Fig. \ref{figpp}a shows that for the smallest atom considered (i.e. Neon) 
the role of $p$-type orbitals is quite small (1.8\%). 
However, it increases almost linearly with $N_e^{-1/3}$ and
for Radon $\Delta^\mathrm{EXX}$ is already above $8\%$. 
To investigate the semiclassical limit for 
$N_e\rightarrow \infty$ we further consider hydrogenic 
orbitals \cite{lieb}. After some algebra the following relations are obtained:
\begin{equation}
\tau^W[\dens_{n0}](0)=\frac{Z^5}{\pi n^3}\;\ ; \;\ \tau^W[\dens_{n1}](0)=\frac{(n^2-1)Z^5}{3\pi n^5} \, ,
\end{equation}
\begin{equation}
\Delta^\mathrm{HYD}[N_e]=\frac{\sum_{n=2}^{N}(\frac{1}{n^3}-\frac{1}{n^5}) }{\sum_{n=1}^{N} \frac{1}{ n^3} +  \sum_{n=2}^{N}   (\frac{1}{n^3}-\frac{1}{n^5}) } \, ,
\label{eq:deltahyd}
\end{equation}
where $N$  is the number of filled shells ($N=N_s=N_p$ for noble atoms) and the total number of electron is
$N_e=(N^3+6N^2+11N-6)/6$ if $N$ is odd and
$N_e=(N^3+6N^2+14N)/6$ otherwise.
A plot of $\Delta^\mathrm{HYD}$ as a function of $N_e^{-1/3}$ 
is reported  Fig. \ref{figpp}a.
For small atoms  $\Delta^\mathrm{HYD}$ differs quite significantly
from $\Delta^\mathrm{EXX}$ because, due to screening effects,
EXX orbitals are rather different from hydrogenic orbitals.
Nevertheless, differences reduce for larger atoms and both 
quantities converge to similar results for large $N_e$ values.
Using the hydrogenic orbitals the limit $N_e\rightarrow \infty$
can be computed exactly as
\begin{equation}
\lim_{N_e\rightarrow \infty} \Delta^\mathrm{HYD}[N_e] =  \frac{\zeta(3)-\zeta(5)}{2\zeta(3)-\zeta(5)}=0.12078 \, ,
\end{equation}
where $\zeta(x)$ is the Riemann function.
Thus, in the semiclassical limit, the  $p$-type orbitals contribute to $\etau$
at the origin by 12\%, clearly showing that the VW functional 
can be quite inaccurate in the cusp region.

Then we consider positively charged noble atoms. 
In Fig. \ref{figpp}b we report $\Delta^\mathrm{EXX}$ 
for Ne and Ar isoelectronic series, as a function of $Z^{-1/3}$. 
Also in these cases, in the limit for  $Z\rightarrow\infty$ the hydrogenic orbitals model becomes exact 
as there are no screening effects on the nuclear charge for any shell.
Thus the $Z\rightarrow\infty$ limit can be obtained from Eq. (\ref{eq:deltahyd}): for Neon ($N=2$, $N_e=10$) we obtain  $\Delta^\mathrm{HYD}=1/13$,
for Argon ($N=3$, $N_e=18$)  $\Delta^\mathrm{HYD}\approx 0.983$. 
These values are shown as horizontal lines in  Fig. \ref{figpp}b, 
and are correctly approached by  calculated values of $\Delta^\mathrm{EXX}$. 



We now turn to analyze Eq. (\ref{eq5}) in the limit $r\rightarrow\infty$.
In the asymptotic region only the outer shell (with  $n=\tilde{n}$ and $l=\tilde{l}$) 
contributes to the density \cite{homodens}, 
i.e. $\dens\rightarrow \adens=\dens_{\tilde{n}\tilde{l}}$ (with a $\tilde{}\;$ we indicate 
 asymptotic quantities).
From  Eq. (\ref{eq5}) the asymptotic expression for $\taup$ is obtained:
\begin{equation}
\label{eq6}
\taup=\etau -\tau^{W} 
\underset{r\rightarrow \infty}{\longrightarrow}
\etau_{\tilde{n}\tilde{l}}-\tilde{\tau}^W=
\frac{\tilde{l}(\tilde{l}+1)}{2} 
\frac{\adens}{r^2}=\ataup \,  .
\end{equation}
where we used that $\tau^{W}\rightarrow\tilde{\tau}^W={\tau}^W[\adens]$.
When $\tilde{l}=0$, i.e. for $s$-type 
outer-valence electrons, Eq. (\ref{eq6}) indicates
that $\etau$ asymptotically approaches $\tau^W$, exactly \cite{dftbook2,burke,alva,vitos,ernz00,llMGGA}.
On the other hand, 
for systems with degenerate 
outer shells ($p$-, $d$-, $\ldots$,-type
electrons), e.g.  all atoms but groups I or II, 
$\ataup$ is not zero and decays with the same exponential decay of the density: thus it is very important 
in the near and middle asymptotic region \cite{suppinfo}. 
Eq. (\ref{eq6}) is asymptotically exact, in the sense that no additional terms with different radial powers are present, and  
it is valid for any spherical system, clarifying and generalizing previous observations \cite{ernzer,elf,johnson09}.


We then consider the following exchange energy density per particle 
\begin{equation}\label{eq2}
\epsilon_x = \epsilon_x^{LDA} F_x \;\; \mathrm{with}\;\; F_x = A \frac{8\pi}{\sqrt{15}} \sqrt{\alpha} \, ,
\end{equation}
where $\epsilon_x^{LDA}=-C_x\dens^{1/3}$ (with $C_x=(3/4)(3/\pi)^{1/3}$) 
is the exchange energy per particle in the local density approximation,
$F_x$ is the exchange enhancement factor and  
$\alpha=(\etau-\tau^W)/\tautf$ is the
well known meta-GGA ingredient \cite{TPSS},
with $\tautf=C_s\dens^{5/3}$  (with $C_s=(3/10)(3\pi^2)^{2/3}$) being the 
Thomas-Fermi KE density \cite{TF}; $A$ is a positive constant (with the factor $8\pi/\sqrt{15}=\sqrt{8C_s}/C_x$ included 
for simplicity; see below).
The expression in Eq. (\ref{eq2}), i.e. an enhancement factor ($F_x$) 
proportional to $\sqrt{\alpha}$ is not new.
It was derived in Ref. \cite{luc2d} with $A\approx 0.3$ to describe 
exactly the exchange in the large-gradient limit of quasi-two dimensional (quasi-2D) systems.
Previous investigations of $\alpha$ were instead mostly related to the 
description of bonds \cite{johnson09}.
Here we consider a completely different physics and study the tail region of atoms.

We rewrite Eq. (\ref{eq2}) as 

\begin{equation}
\epsilon_x=-A C_x \frac{8\pi}{\sqrt{15}} \sqrt{\frac{\etau-(\nabla \dens)^2/(8\dens)}{\dens\,  C_s}}=(-A)\frac{\Theta}{\dens} \, ,
\end{equation}
with $\Theta=\sqrt{8\etau \dens-(\nabla \dens)^2}$.
From  Eq. (\ref{eq6}) we have that
$\Theta\rightarrow 2\sqrt{\tilde{l}(\tilde{l}+1)} (\adens/r)$.
Thus
\begin{equation} \label{eqeps}
\epsilon_x \rightarrow 2 (-A)\sqrt{\tilde{l}(\tilde{l}+1)}\frac{1}{r} \, ,
\end{equation}
and the the exact asymptotic decay ($\epsilon_x\rightarrow -1/(2r)$) is 
obtained if 
\begin{equation}
A=A'=1/\left( 4\sqrt{\tilde{l}(\tilde{l}+1)}\right) \label{a1eq} \, .
\end{equation}

Concerning the exchange potential, we consider the generalized Kohn-Sham framework 
to write \cite{pot2}
\begin{eqnarray}
\displaystyle v_x \phi_i &=&
[\frac{\partial(\dens\epsilon_x)}{\partial \dens}-\nabla 
\frac{\partial(\dens\epsilon_x)}{\partial \nabla \dens }
]\phi_i
-\frac{1}{2}\nabla(\frac{\partial(\dens\epsilon_x)}{\partial\etau})\nabla\phi_i\nonumber\\
& -& \frac{1}{2}\frac{\partial(\dens\epsilon_x)}{\partial\etau}\nabla^2\phi_i  \nonumber \\
&=& \frac{\partial (\dens \epsilon_x)}{\partial \dens}\phi_i
 -\nabla\cdot  \left [ \frac{\partial (\dens \epsilon_x)}{\partial \nabla \dens} \phi_i 
+\frac{1}{2} \frac{\partial (\dens \epsilon_x)}{\partial \etau} \nabla\phi_i 
\right ] \nonumber  \\
&+&  \left( 
 \frac{\partial (\dens \epsilon_x)}{\partial  \nabla \dens }
\right)\cdot\nabla\phi_i  \, . 
\label{eq4}
\end{eqnarray}
As first step we compute the partial derivatives of 
$\dens\epsilon_x=(-A)\Theta$ 
with respect $\dens$, $\nabla \dens$, $\etau$:
\begin{eqnarray}
\frac{\partial (\dens \epsilon_x)}{\partial \dens} &=& (-A) \frac{4\etau}{\Theta}   
 = (-A) \frac{\Theta}{2\dens} +  \frac{\partial (\dens \epsilon_x)}{\partial \etau} \frac{(\nabla \dens)}{4\dens}\frac{(\nabla \dens)}{2\dens}  \nonumber \\
\frac{\partial (\dens \epsilon_x)}{\partial \nabla \dens} &=& (-A) \frac{(\nabla \dens)}{\Theta} 
=  \frac{\partial (\dens \epsilon_x)}{\partial \etau}   \frac{(\nabla \dens)}{4\dens}  \nonumber \\
\frac{\partial (\dens \epsilon_x)}{\partial \etau}   
 &=& (-A) \frac{4 \dens}{\Theta} \label{dtau} \, .
\end{eqnarray}
Interestingly Eq. (\ref{dtau}) shows 
that the exchange energy per
particle of Eq. (\ref{eq2}) possesses the 
desirable property that $\partial(\dens\epsilon_{x})/\partial\etau< 0$,
which was shown to be essential for an accurate 
description of the optical properties of 
semiconductors at the meta-GGA level \cite{Vignale}. 
In contrast, no other non-empirical meta-GGA functional
recovers this condition.
Inserting the derivatives into Eq. (\ref{eq4}) we find
\begin{eqnarray}
v_x\phi_i 
&=&
(-A)\frac{\Theta}{2\dens}\phi_i 
-\nabla\cdot  \left
[ \frac{\partial (\dens \epsilon_x)}{\partial \etau} \left \{ - \frac{(\nabla \dens)}{4\dens}\phi_i
+\frac{1}{2} \nabla\phi_i \right \}  \right ] \nonumber \\
&+&
\frac{\partial (\dens \epsilon_x)}{\partial \etau}   \frac{(\nabla \dens)}{4\dens}
\left \{ \frac{(\nabla \dens)}{2 \dens}\phi_i
- \nabla\phi_i \right \} \label{eqvvv} \, .
\end{eqnarray}
This expression clearly depends on the considered orbital $\phi_i$.
For the asymptotic properties we have to consider the highest occupied orbital $\phi_i=\phi_H$:
in this case the asymptotic density is $\adens=f\phi_H^2$ and 
the terms in curly braces in Eq. (\ref{eqvvv}) vanish identically.
Thus, we finally obtain 
\begin{equation}
v_x \phi_H = (-A) \frac{\theta}{2 \dens}\phi_H \rightarrow 
 A\sqrt{\tilde{l}(\tilde{l}+1)}\frac{1}{r}\phi_H \label{eqvx} \, , 
\end{equation}
that recovers the exact asymptotic decay ($v_x\rightarrow -1/r$) if 
\begin{equation}
A=A''=1/\left( \sqrt{\tilde{l}(\tilde{l}+1)} \right)  \, . \label{a2eq}
\end{equation}

Despite the two constant $A'$ and $A''$ in Eqs. (\ref{a1eq}) and 
(\ref{a2eq}) differ by a factor of 4, the here proposed expression for $\epsilon_x$
(Eq. (\ref{eq2})) yields asymptotic properties proportional to 
the exact ones for \emph{both} the exchange energy density and the exchange potential.
This is a strong improvement with respect to current meta-GGAs 
where $\epsilon_x$ and $v_x$ decay exponentially and GGA functionals
where either $\epsilon_x$ or $v_x$ can have the exact properties. 

\begin{figure}
\includegraphics[width=\columnwidth]{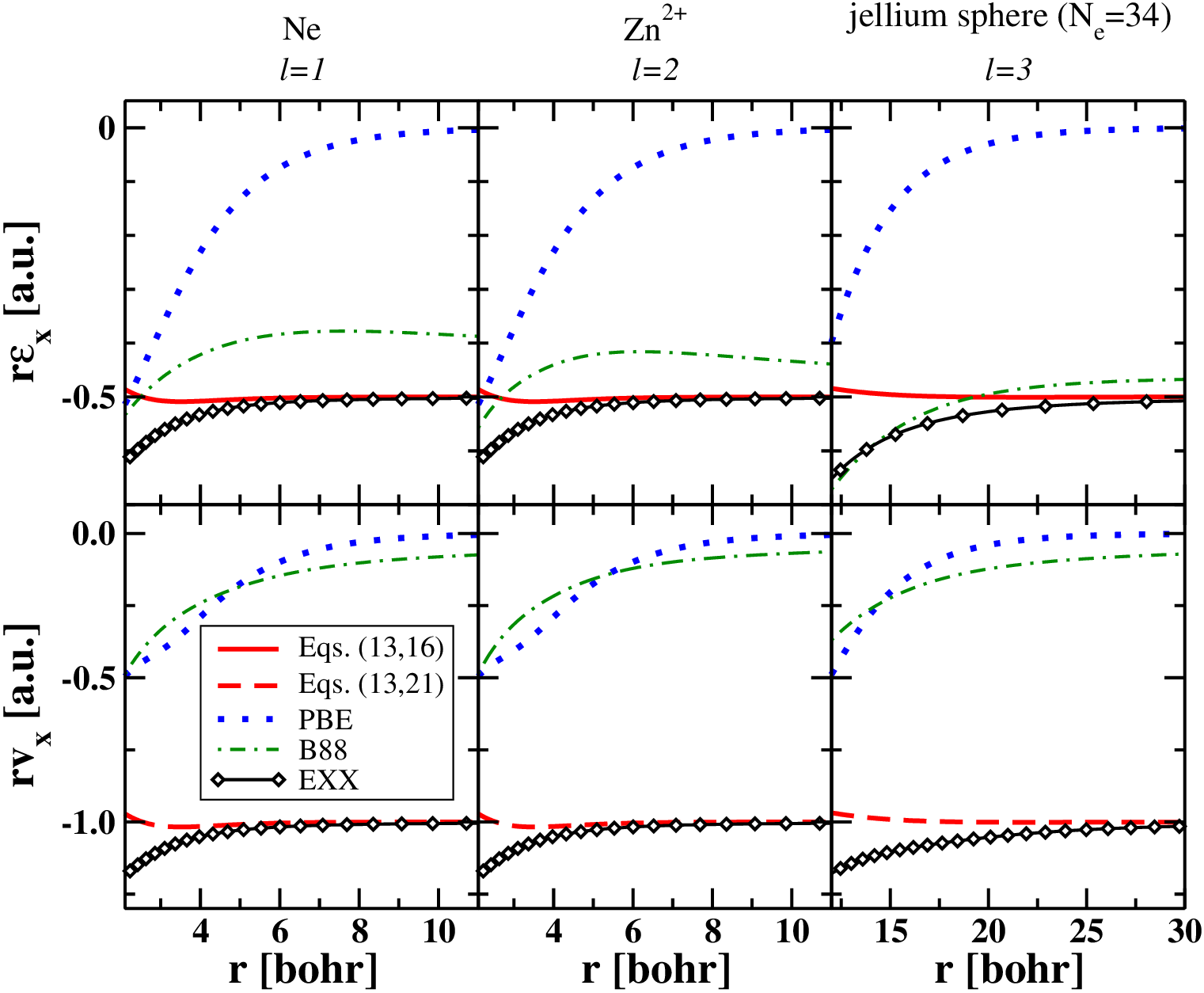}
\caption{Exchange energy per particle $\epsilon_x$ (upper panels) 
and exchange potential $v_x$ (lower panels), multiplied by the radial distance $r$, from different approaches and
for Ne, Zn$^{2+}$, and a jellium sphere with 34 electrons, having $p-$, $d-$, and $f$-type outer shell, respectively.
}
\label{f1}
\end{figure}
To validate  the previous analytic results
we consider in Fig. \ref{f1} the comparison of $\epsilon_x$ and $v_x$
obtained from Eq. (\ref{eq2}) for $p$- and $d$-type 
closed-shell atoms (i.e. Ne and Zn$^{2+}$ ) and for a jellium sphere with 34 electrons and $r_s$=2.07 (having a $f$-type outer shell). 
Results from Eq. (\ref{eq2}) are compared to EXX  and standard GGAs. 

The numerical results confirm that,
unlike other semilocal approximations (e.g. PBE \cite{PBE}),
the simple expression in Eq. (\ref{eq2}) yields the correct
asymptotic behavior for $\epsilon_x$ (i.e. $-1/(2r)$).
Fig. \ref{f1} also shows that Eq. (\ref{eq2}) is also in better agreement with EXX than the B88 exchange \cite{becke}; the latter, in fact, 
approaches  $-1/(2r)$ only at very large distances.
For the exchange potential  Eq. (\ref{eq2}) and Eq. (\ref{a2eq}) give a $-1/r$ behavior, whereas PBE and B88 decay much 
faster (exponentially and proportional to $-1/r^2$, respectively). 

We thus propose that the expression in Eq. (\ref{eq2}) 
can be a powerful tool in the development of 
meta-GGA exchange functionals. Clearly completely new functional forms need
to be developed in order to satisfy  Eq. (\ref{eq2}) in the asymptotic region.
Moreover, additional work needs to be done  to take into account the 
$l$-dependence in Eqs. (\ref{a1eq}) and (\ref{a2eq}) as well as 
investigations for molecules, where the anisotropy of $\etau$ can also play 
a role \cite{pccp14} and the KS exchange potential can show asymptotic barrier-well structures \cite{prlasy}. 

Finally, Eqs. (\ref{eq6}) can be used to derive an asymptotically correct 
expression for the non-interacting KE at the GGA level of theory, i.e. 
$\tau=\tautf F_s(s)$, where $F_s(s)$ is the KE enhancement factor which depends
on the reduced gradient $s=|\nabla\dens|/(2(3\pi^2)^{1/3}\dens^{4/3})$.
In fact, recalling that for an exponential density the expression $\dens\epsilon_x^{LDA}(4\pi/9)(s/\ln(s))$ decays as $-\dens/(2r)$ \cite{xcrev,becke} 
then $(20/27)\tautf(s/(\ln(s))^2$ will decay as $-\dens/(2r^2)$.
Thus
\begin{eqnarray}
F_s
&\rightarrow& \frac{5}{3}s^2 
\left(
 1 + \frac{4}{9}   
\frac{\tilde{l}(\tilde{l}+1) }{\ln^2(s)} 
\right) 
\, ,
\label{e9}
\end{eqnarray}
represents a new exact constraint for GGA kinetic energy functionals in the asymptotic region. 

In conclusion, from a  simple analytic expression (Eq. (\ref{eq5})) we have shown that:
i)
$p$-type electrons largely contribute to the positive-defined 
KE density at the nucleus. This property has been derived considering
the linearity of the VW functional at the cusp.
For the Neon atom (with only one $p$-shell) in the limit of infinite nuclear charge, $\Delta\approx 8\%$; 
in the semiclassical limit of neutral large atoms, $\Delta$ reaches $12\%$. 
Thus the physics of the KE density near
the nucleus region has fully non-local features (i.e. $\dens_{p}$ is a non-local functional of $\dens$), which 
can be hardly captured by semilocal ingredients. 

ii) The asymptotic expression of the Pauli excess KE density has been used to construct a
new meta-GGA exchange functional: a simple enhancement factor term proportional to 
$\sqrt{\alpha}$ can well describe the asymptotic behavior of $\epsilon_x$ and $v_x$, 
as well as the quasi-2D density regime.
This is an important achievement of 
the meta-GGA level of theory, with respect to the GGA 
one. In fact, at the GGA level, the infinite barrier model 
quasi-2D limit can be described only if the GGA exchange enhancement 
factor decays as $F_x\rightarrow s^{-1/2}$ \cite{q2D}, the asymptotic 
behavior of $\epsilon_x$ can be described 
only if $F_x\rightarrow s/\ln(s)$ \cite{becke}, 
while the asymptotic behavior of $v_x$ can be described only if $F_x$ 
diverges at least as $s$ \cite{arm13}. 
Thus, important future developments in the construction
of accurate non-empirical meta-GGA DFT functionals can be foreseen.
Nevertheless, we remark that 
the inclusion of the present results
into a practical tool still requires additional effort,
in particular for the development of an appropriate
and flexible enough meta-GGA functional 
form in order to correctly describe asymptotic and quasi-2D regimes. 
Finally, Eq. \ref{e9}  is also relevant for future development of approximated KE functionals 
with improved asymptotic behavior.

\end{document}